\documentclass[a4paper,twocolumn,fleqn]{article}

\usepackage{amsmath}               
\usepackage{helvet}
\usepackage{graphicx}
\usepackage[dvips]{color}
\usepackage{jpfr}                  

\newcommand{\phiG}{\phi_{\scriptscriptstyle G}}
\newcommand{\Mdota}{\dot M_{\rm a}}
\newcommand{\Mdotw}{\dot M_{\rm w}}

\newcommand{\Bvec}{\mathbf B}

\newcommand{\nablavec}{\mathbf \nabla}
\newcommand{\vvec}{\mathbf v}

\newcommand{\dr}{{\rm d}r}
\newcommand{\ri}{r_{\rm i}}

\newcommand{\tv}{t^{\rm v}}
\newcommand{\tRij}{t^{\scriptscriptstyle \rm R}_{ij}}
\newcommand{\tBij}{t^{\scriptscriptstyle \rm B}_{ij}}

\begin{document}

\title{Enhanced MHD Transport in Astrophysical Accretion Flows: Turbulence, Winds and Jets}

\author{Peter~B.~DOBBIE\sup{1}, Zdenka~KUNCIC\sup{1}, Geoffrey~V.~BICKNELL\sup{2} and Raquel~SALMERON\sup{2}}

\affiliation{
  \sup{1}School of Physics, University of Sydney, NSW 2006, Australia \\
  \sup{2}Research Centre for Astronomy and Astrophysics, Australian National University, ACT 2611, Australia}

\date{\ \ }

\email{p.dobbie@physics.usyd.edu.au}

\begin{abstract}
Astrophysical accretion is arguably the most prevalent physical process in the Universe; it occurs during the birth and death of individual stars and plays a pivotal role in the evolution of entire galaxies. Accretion onto a black hole, in particular, is also the most efficient mechanism known in nature, converting up to 40\% of accreting rest mass energy into spectacular forms such as high-energy (X-ray and gamma-ray) emission and relativistic jets. Whilst magnetic fields are thought to be ultimately responsible for these phenomena, our understanding of the microphysics of MHD turbulence in accretion flows as well as large-scale MHD outflows remains far from complete. We present a new theoretical model for astrophysical disk accretion which considers enhanced vertical transport of momentum and energy by MHD winds and jets, as well as transport resulting from MHD turbulence. We also describe new global, 3D simulations that we are currently developing to investigate the extent to which non-ideal MHD effects may explain how small-scale, turbulent fields (generated by the magnetorotational instability -- MRI) might evolve into large-scale, ordered fields that produce a magnetized corona and/or jets where the highest energy phenomena necessarily originate.
\end{abstract}


\maketitle


\section{\label{sec:1}Introduction}
It is widely accepted that high-energy astrophysical sources such as active galactic nuclei (AGN), gamma-ray bursts and some X-ray binaries are powered by accretion of matter onto a central black hole.  Since the standard theory of astrophysical disk accretion was formulated over 30 years ago \cite{sha73,novthorn73}, arguably the most important advance in our understanding of the process by which matter in the disk can shed its angular momentum and release its gravitational binding energy has come from computational modelling. Numerical simulations demonstrate unequivocally that the magnetorotational instability (MRI, \cite{vel59,cha60,bal91,bal98}) can produce magnetohydrodynamic (MHD) turbulence and enhanced angular momentum transport (see \cite{bal03} for a review).  The presence of even a very weak magnetic field is the key ingredient: it completely changes the dynamics from a keplerian flow which is hydrodynamically stable even at high Reynolds numbers (as recently verified experimentally \cite{ji06}) to one which is unstable to the rapid growth of MHD modes leading to turbulence in the nonlinear regime.

It is over 20 years since the first MHD simulations of astrophysical accretion flows were carried out \cite{uch85,shi86}.  Notwithstanding the important advances made to date \cite{bal91,haw01,ste01,sto01,haw02,stepap02n,katmin04n,kig05,mck06n,macmat08n}, numerical simulations have so far been unable to resolve two key outstanding issues:
\begin{enumerate}
\item How are the high rates of mass accretion inferred in the most powerful sources achieved?
\item How are the outflows and jets observed across the mass spectrum of accreting sources produced?
\end{enumerate}
In what follows, we briefly address each of these open questions and suggest how they may be connected and mutually resolved by a generalized model for MHD disk accretion.

\begin{figure}[ht]
\centerline{\includegraphics[width=8cm,clip]{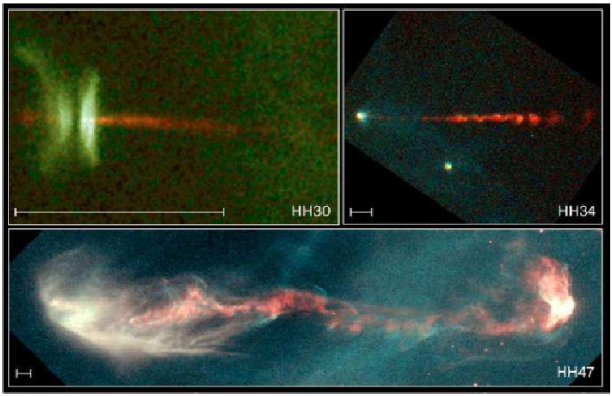}}
  \caption{Hubble Space Telescope  optical images of jets in  young stellar objects. Photo credits: C. Burrows, J. Morse (STScI), J. Hester (AZ State U.), NASA.}
  \label{fig:jets}
\end{figure}

The most powerful accreting sources (i.e. quasars and other active galaxies) are fuelled by accretion onto a supermassive ($10^{6-9}\,M_\odot$) black hole. They produce radiative luminosities that can exceed those of normal galaxies by several orders of magnitude (e.g. up to $10^{48} \rm erg \, s^{-1}$), indicating mass accretion rates which can exceed 100 solar masses per year ($1 \, M_\odot \rm yr^{-1} \approx 6\times10^{25}\,\text{g}\,\text{s}^{-1}$). This is strictly a lower limit because accretion can also drive mechanical outflows, in some cases with inferred kinetic powers that are considerably greater than the observed radiative luminosity (e.g. the famous M87 jet -- see \cite{jolkun07} and references therein). Indeed, the fact that collimated jets are observed across a wide range of accreting sources (see \cite{livio99} for a review), including those that are non-relativistic (e.g. protoplanetary systems, young stellar objects, see Fig.~\ref{fig:jets}, and neutron stars, see especially \cite{fen04}), suggests that accretion provides an effective, generic mechanism for powering jets and other energetic plasma outflow phenomena.

Magnetic fields are required to collimate jet plasma and to account for the observed radio synchrotron jet emission. MHD simulations have revealed that a net poloidal component of the magnetic field, $B_z$, is required to produce winds and jets from an accretion disk (see \cite{pud06} for a review). Interestingly,  high mass accretion rates can only be  achieved in the simulations if a large-scale MHD outflow is present (e.g. \cite{ste01,kig05}). It is particularly noteworthy that comparably high accretion rates cannot be achieved through MRI-driven MHD turbulence alone, despite the fact that the effective turbulent viscosity is ``enhanced'' relative to kinematic fluid shear viscosity (e.g. \cite{sto01,haw01,haw02}). Therefore, magnetized jets (and by implication, MHD outflows in general) must be primarily responsible for truly enhanced transport in astrophysical accretion flows. This conclusion is consistent with early analytical models  \cite{bla82,pud83}  proposing that  accretion is facilitated by the vertical transport of angular momentum resulting from MHD torques on the disk surface. It is also consistent with non-ideal MHD simulations indicating that radial transport of angular momentum by turbulent stresses may be restricted to interior regions near the disk midplane only \cite{sal07}.
 
Although MHD simulations indicate that a poloidal field component is essential for launching accretion disk jets \cite{beck08}, it is not yet clear how such a field component arises in the case of accretion onto black holes, which have no intrinsic magnetic moment.  The magnetic flux in the disk must originate from the random field in the interstellar medium (or from the companion star in the case of a black hole X-ray binary). This field is then amplified into a dominant toroidal configuration in the disk by the MRI.  Simulations have yet to demonstrate the feasibility of creating large-scale fields via an inverse cascade process involving either stochastic reconnection of turbulent fields \cite{chris01} or reconnection of buoyant flux loops emerging from the disk surface \cite{touprin96}.  Models which require {\it a priori} large-scale flux loops \cite{lovrom95n,hayshi96n,romust98n}, a poloidal B field (e.g. \cite{katmin04n}) or a spinning black hole (\cite{mck06n,mckbla09n}) to produce jets are too restrictive to explain the observed ubiquity of jets and outflows across the wide range of accreting systems. One of the most interesting numerical results to date is that of Machida and Matsumoto \cite{macmat08n}.  Their global 3D simulations show the evolution of a large-scale poloidal field from an initially weak toroidal field. However, no simulation to date has shown an initial random field in a generic accreting system evolve into a configuration favourable for jet production. Similarly, while Machida {\it et al.} \cite{macnak06n} have studied the effect of radiative cooling on optically \emph{thin} black hole accretion flows, no simulations to date have shown the effect of optically thick radiative cooling on the global, 3D evolution of the magnetic fields.

We are developing new global 3D MHD simulations to directly confront this challenging problem and obtain new insights into the evolution of magnetic field topology in black hole accretion flows. We aim to test our hypothesis that  turbulent reconnection and resistive dissipation, as well as radiative losses by the plasma, play pivotal roles in the evolution and steady-state properties of MHD accretion flows. We suggest that the microphysics of MHD accretion can govern large-scale, macrophysical phenomena that ultimately determine the observational appearance and hence, classification, of accreting black hole sources.

Numerical approaches to date have been limited by one or more of the following drawbacks: the use of the shearing box approximation (see, e.g., \cite{reg08,bod08} for some limitations of this approach); use of a non-conservative numerical scheme (e.g. \cite{hay06}) resulting in unphysical levels of numerical dissipation which prevent a quantitative analysis of energy transport; neglect of finite resistivity making it difficult to realistically model magnetic reconnection; and neglect of radiation, resulting in the unphysical situation where heat generated in the disk at the end of a turbulent cascade cannot be radiated away, so the disk puffs up. This also makes it impossible to compare model results against observations of luminosity and spectrum.

To implement our model, we are using and extending FLASH\footnote{FLASH is freely available at http://flash.uchicago.edu.} \cite{fry00}, the public MHD code developed at the University of Chicago.  FLASH provides the following features which make it well suited to our model: it implements adaptive mesh refinement (AMR); it solves the equations of MHD in conservation form thus explicitly conserving energy;
it uses a modified piecewise-parabolic method (PPM \cite{woo84,col84}) which is significantly more
accurate than some other widely used codes (e.g. \cite{hay06});
it uses the constrained transport method \cite{eva88} to enforce divergence free magnetic fields; and it is modular and extensible, allowing us to develop a radiation MHD module.  As such it represents the next generation of MHD codes.

\begin{figure*}[t]
\centering
  \includegraphics[width=0.9\textwidth]{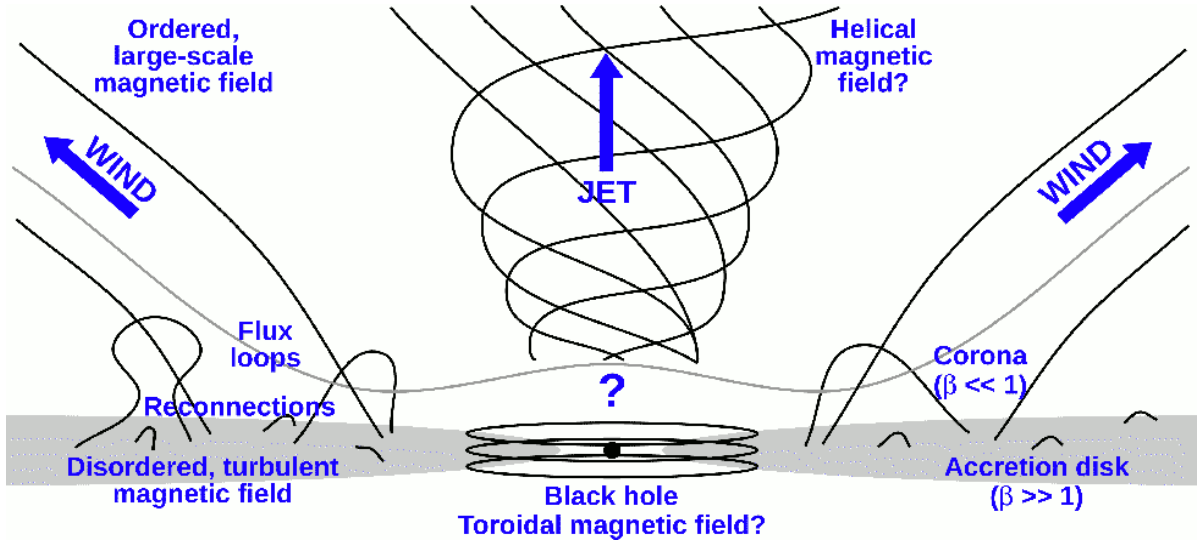}
  \caption{A schematic illustration of the inner regions of an MHD accretion flow around a black hole. $\beta$ is the usual plasma beta (the ratio of gas to magnetic pressures). The wind can drive mass loss from the disk, while the jet may be dominated by Poynting flux.}
  \label{fig:bhaccretion}
\end{figure*}

The organization of this paper is as follows. In  \S2, we briefly review an analytic model  for turbulent MHD black hole disk accretion that forms the theoretical basis for our numerical investigations. In \S3, we describe the  simulations we are currently developing, including non-ideal MHD effects and radiation. We present some concluding remarks in \S4.

\section{\label{sec:2}MHD Disk Accretion Theory}
The analytical foundation for our approach is based on the model of Kuncic and Bicknell \cite{kun04}. This model employs a mass-weighted statistical averaging of the MHD equations to obtain a mean-field description of  turbulent MHD disk accretion that is steady-state and axisymmetric in the mean. Angular momentum and energy are transported radially outwards by turbulent Maxwell stresses and vertically outwards by a large-scale MHD wind and/or jet. The inner region of the black hole accretion disk is also surrounded by and magnetically coupled to a hot, diffuse corona, analogous to the solar corona.

Two important observational predictions of the model are: 1. The disk emission spectrum is degraded by the electromagnetic extraction of gravitational binding energy from the accreting matter (see \cite{kunbick07a,kunbick07b}); and 2. The presence of a magnetized jet substantially enhances the rate of mass accretion in the disk and hence, the rate of black hole growth, resulting in a correlation between black hole mass and radio emission (see \cite{jolkun08}).
The model is schematically illustrated in Fig.~\ref{fig:bhaccretion}.
The main results pertinent to our simulations are summarised below -- more details can be found in the original paper \cite{kun04}.

\subsection{Statistical Averaging}
In the mass-weighted statistical averaging approach, all variables are decomposed into mean and fluctuating parts, with intensive variables such as velocity $\vvec$ mass averaged according to
\begin{equation}
v_i = \tilde{v}_i + v^{\prime}_i, \enspace \langle\rho v^{\prime}_i\rangle = 0 \qquad,
\end{equation}
while extensive variables such as density $\rho$, pressure $p$ and magnetic field $\Bvec$, are averaged the following way:
\begin{align}
\rho = \bar{\rho} + \rho^{\prime}, \enspace &\langle \rho^{\prime}\rangle = 0 \qquad ,\\
p = \bar{p} + p^{\prime}, \enspace &\langle p^{\prime}\rangle = 0 \qquad ,\\
B_i = \bar{B}_i + B^{\prime}_i, \enspace &\langle B^{\prime}_i\rangle = 0 \qquad.
\end{align}

For clarity, in the following equations the tilde and bar have been omitted from the averaged intensive and extensive variables, respectively: averaged quantities are implicitly assumed. We consider below only the simplest case where $\bar B_i =0$, although it is straightforward to generalize to the case with a nonzero net mean magnetic field.  We have also omitted negligible correlation terms.\footnote{In particular, triple correlation terms of the form $\langle t_{ij}v^{\prime}_j \rangle$ are assumed negligible compared to analogous correlations with the mean fluid velocity $\langle t_{ij} \rangle\tilde{v}^{\prime}_j$.} 

\subsection{Mass Transfer}
Integration of the mean-field continuity equation gives
\begin{equation}
\Mdota (r) + \Mdotw (r) = \text{constant} = \dot{M} \qquad ,
\end{equation}
where $\Mdota (r)$ is the mass accretion rate and $\Mdotw (r)$ is the mass  outflow rate associated with a mean vertical velocity at the disk surface, i.e. at the base of a disk wind.
Under steady-state conditions, the radial mass inflow decreases towards small $r$ at the same rate as the vertical mass outflow increases in order to maintain a constant net mass flux, $\dot{M}$, which is the net accretion rate at $r = \infty$.

\subsection{Momentum Transfer}
Statistical averaging of the momentum equation yields
\begin{align}
\notag\frac{\partial(\rho v_i)}{\partial t} &+ \frac{\partial(\rho v_i v_j)}{\partial x_j} \\
&= -\rho\frac{\partial \phiG}{\partial x_i} - \frac{\partial p}{\partial x_i} + \frac{\partial}{\partial x_j}\left( \tRij + \langle \tBij \rangle \right) \; ,
\end{align}
where $\phiG = - GM ( r^2 + z^2)^{-1/2}$ is the gravitational potential of the central mass $M$,
$\tRij = - \langle \rho v^{\prime}_i v^{\prime}_j \rangle$
is the Reynolds stress, and the turbulent Maxwell stress is
\begin{equation}
\langle \tBij \rangle = \langle \frac{B_i B_j}{4\pi} \rangle - \delta_{ij}\langle \frac{B^2}{8\pi} \rangle \; .
\end{equation}

Integration of the azimuthal component of the momentum equation yields
\begin{align}
\notag\Mdota v_\phi r &- \Mdota(\ri) v_\phi(\ri) \ri \\
\notag= &-2 \pi r^2 T_{r\phi} + 2 \pi \ri^2 T_{r\phi}(\ri) \\
&+ \int^r_{\ri}\left[v_\phi r \frac{\text{d}\Mdota}{\text{d}r} - 4 \pi r^2 \langle t_{\phi z}\rangle^+\right] \dr \; ,
\label{eq:1}
\end{align}
where quantities calculated at the disk surface
are denoted by a `$+$' superscript,
$\ri$ denotes the radius at the innermost stable circular orbit and
\begin{equation}
T_{r\phi} = \int^{+h}_{-h} \langle t_{r\phi} \rangle \text{d}z
\end{equation}
is the $r\phi$ component of the Maxwell stress integrated over the vertical scaleheight $h$.

The left hand side of eqn.~(\ref{eq:1}) describes the change in angular momentum flux associated with inflow from an outer radius $r$ to $\ri$. The first two terms on the right hand side describe the rate of radial transport of angular momentum due to MHD stresses in the disk.  The terms in the integrand on the right hand side describe the {\it vertical} transport of angular momentum resulting from mass loss in a wind and an MHD torque on the disk surface, respectively. These effects are not modelled in standard accretion disk theory.
In summary, the model we consider includes contributions from both radial and vertical transport of angular momentum to the overall mass accretion rate: angular momentum is transported radially outwards by internal MHD stresses and vertically outwards by both mass outflows and MHD stresses acting over the disk surface.

Previous simulations \cite{ste01,kig05,sal07} as well as semi-analytic models (e.g., \cite{cam03}) show that in the presence of a large-scale, open, mean magnetic field, angular momentum is transported at small disk radii more efficiently in the vertical direction by large-scale magnetic torques (Poynting flux) than radially by MHD turbulence.  Our simulations will compare the contribution from these processes as well as from the vertical transport of angular momentum by mass outflows.

\subsection{Energy Transfer}
Accretion extracts gravitational binding energy from the accreting matter and converts it into mechanical (e.g. kinetic, Poynting flux) and non-mechanical (e.g. radiative) forms.  The rate at which this occurs is determined by the keplerian shear in the bulk flow, $s_{r\phi} = \frac{1}{2}r \partial\Omega/\partial r$, with $\partial\Omega/\partial r = -\frac{3}{2}\Omega / r$.  

The rate per unit disk surface area at which energy is emitted in  the form of electromagentic radiation is determined by the internal energy equation:
\begin{align}
\notag\frac{\partial u}{\partial t} + &\frac{\partial}{\partial x_i}\left(uv_i + \langle uv^{\prime}_i \rangle\right) \\
\notag\approx &-pv_{i, i} - \langle pv^{\prime}_{i, i} \rangle - \langle F_{i, i} \rangle \\
&+ \langle\frac{J^2}{\sigma}\rangle + \langle \tv_{ij}v^{\prime}_{i, j} \rangle \; ,
\label{eq:3}
\end{align}
where $u$ is the gas plus radiation energy density, $\mathbf F$ is the radiative flux, $\mathbf J$ is the current density, and  $\tv_{ij}$ is the viscous stress tensor.  The terms on the left hand side describe the total rate of change of gas plus radiation energy density and the terms on the right hand side describe work done by compression in the flow against the gas and radiation pressure, radiative losses, mean field ohmic heating, and viscous dissipation (heating).  The last term requires some comment since the molecular viscosity in the {\it mean flow} is generally considered negligible in accretion disks.  However, at the high-wavenumber end of a turbulent cascade, it can become important in converting the turbulent energy into heat.  The viscous stress tensor is
\begin{equation}
\tv_{ij} = 2\nu\rho s_{ij} \qquad ,
\end{equation}
where $\nu$ is the coefficient of kinematic shear viscosity and $s_{ij}$ is the shear tensor:
\begin{equation}
s_{ij} = \frac{1}{2}\left( v_{i, j} + v_{j, i} - \frac{2}{3}\delta_{ij}v_{k, k} \right) \; .
\end{equation}

The source terms determine the rate at which energy is converted into random particle energy (some of which is then converted into radiation) and into bulk kinetic energy.  If there are negligible changes in the internal energy of the gas and the turbulent energy is dissipated at the end of a turbulent cascade at a rate equivalent to its production, then eqn.~(\ref{eq:3}) implies that the disk radiative flux emerging from the disk surface is
\begin{equation}
F^+_{\rm d} \approx \frac{1}{2}T_{r\phi}r\frac{\partial \Omega}{\partial r} = -\frac{3}{4}T_{r\phi}\Omega \; .
\label{eq:2}
\end{equation}

The level of these turbulent MHD stresses available to dissipate the internal energy in turn depends on how efficiently the gravitational binding energy extracted by accretion is converted into other forms (both mechanical and non-mechanical).  That is, $T_{r\phi}$ is determined by the angular momentum conservation equation (c.f. \ref{eq:1}):
\begin{align}
\notag-T_{r\phi}(r) &= \frac{\Mdota v_\phi r}{2\pi r^2} \left[ 1 - \frac{\Mdota (\ri)}{\Mdota (r)} \left(\frac{\ri}{r}\right)^{1/2} \right] \\
\notag&- \left(\frac{\ri}{r}\right)^2 T_{r\phi}(\ri) \\
&- \frac{1}{2\pi r^2} \int^r_{\ri}\left[v_\phi r \frac{\text{d}\Mdota}{\text{d}r} - 4 \pi r^2 \langle t_{\phi z}\rangle^+\right] \text{d}r.
\end{align}
Substituting this into (\ref{eq:2}) yields the following more general solution for the disk radiative flux:
\begin{align}
\notag F^+_d(r) &\approx \frac{3GM\dot{M}_a(r)}{8\pi r^3} \left[ 1 - \frac{\dot{M}_a(r_i)}{\dot{M}_a(r)} \left(\frac{r_i}{r}\right)^{1/2} \right] \\
\notag&- \frac{3}{4}\left(\frac{r_i}{r}\right)^2 T_{r\phi}(r_i)\Omega \\
&- \frac{3\Omega}{8\pi r^2} \int^r_{r_i}\left[v_\phi r \frac{\text{d}\dot{M}_a}{\text{d}r} - 4 \pi r^2 \langle t_{\phi z}\rangle^+\right] \text{d}r.
\end{align}

This is the generalized solution for the radiative flux of a turbulent MHD accretion disk.  It can be expressed as
\begin{equation}
F^+_{\rm d} (r) \approx \frac{3GM\Mdota (r)}{8\pi r^3} [f_{\rm a}(r) - f_{\rm w}(r)] \; ,
\label{eq:4}
\end{equation}
where
\begin{equation}
f_{\rm a} (r)  = \left[ 1 - \frac{\Mdota (\ri)}{\Mdota (r)} \left(\frac{\ri}{r}\right)^{1/2} \right] - \frac{2\pi \ri^2 T_{r\phi}(\ri)}{\Mdota (r) r^2\Omega}
\label{eq:5}
\end{equation}
is a dimensionless factor that parameterizes the available accretion energy flux, the last term describing the rate at which MHD stresses at the innermost stable circular orbit locally dissipate turbulent energy, and
\begin{align}
\notag f_{\rm w}(r) = &\frac{1}{\dot{M}_{\rm a}(r)r^2\Omega} \cdot \\
&\int^r_{r_i}\left[v_\phi r \frac{\text{d}\dot{M}_a}{\text{d}r} - 4 \pi r^2 \langle t_{\phi z}\rangle^+\right] \text{d}r
\end{align}
is the fractional rate of vertical energy transport from the disk (the `w' subscript denoting a wind). This is a correction factor which takes into account partitioning of accretion power into  non-radiative forms.

Note the difference between this model and standard disk accretion theory \cite{sha73}.  In the latter, all the gravitational binding energy is locally dissipated and assumed to be converted to radiation: $\text{d}\Mdota /\dr = 0$ and $\langle t_{\phi z} \rangle^{+} = 0$ so that $f_{\rm w} = 0$.
This difference will be manifested by a disk spectrum which differs from that predicted by the standard model since the local disk temperature $T(r)$ is reduced if energy is channelled away by outflows from the disk surface. This will affect the emission spectrum arising from the  innermost regions of the disk, where the temperature is highest and where jets and outflows originate.
Assuming local blackbody emission, the disk luminosity spectrum can be calculated by summing up the contributions from each annulus: 
\begin{equation}
L_{d, \nu} = 2\int^\infty_{r_i} \pi B_{\nu}[T(r)] \, 2\pi r \, \text{d}r \; ,
\end{equation}
where $B_{\nu}$ is the Planck function, $T(r) = [F^+_{\rm d} (r)/\sigma]^{1/4}$ is the effective disk temperature of each annulus and $\sigma$ is the Stefan-Boltzmann constant.  We expect the disk radiative efficiency to be lower than the canonical $\simeq 10\%$ predicted by the standard model when vertical transport of angular momentum is important. The radiative efficiency is given by the ratio of disk luminosity to accretion power: $L_{\rm d} / P_{\rm a}$. The total accretion power is calculated from
\begin{equation}
P_{\rm a} = 2 \int_{\ri}^\infty \frac{3GM\Mdota (r)}{8\pi r^3} f_{\rm a}(r) \, 2\pi r \, \dr \; .
\end{equation}
If there is no wind mass loss from the disk, so that $\Mdota$ is constant, this reduces to the familiar result $P_{\rm a} = \frac{1}{2} GM \Mdota / \ri \approx \frac{1}{12}\Mdota c^2$, in the Newtonian approximation for a nonrotating black hole.

\section{MHD Accretion Simulations}

As described earlier, there is a real need for a new program of MHD simulations to advance our knowledge of accreting black hole systems. Our numerical work is motivated by the following:
\begin{list}{}{\itemsep=3pt\topsep=1pt}
\item[1.] We need to explain the macrophysics of observed phenomena in AGN and other accreting systems, viz., high mass accretion rates and jets/winds, and test the hypothesis that they are related by large-scale MHD processes.
\item[2.] We need to improve our understanding of the microphysics in order to explain how small-scale, local MHD processes can evolve into large-scale, global phenomena.
\item[3.] We need to explicitly calculate the radiation emitted by a black hole accretion disk in order to directly compare against the observational data.
\end{list}

FLASH solves the time-dependent equations of compressible non-ideal MHD. In non-dimensional conservation form these are:
\begin{align}
\frac{\partial \rho}{\partial t} &+ \nablavec \cdot (\rho{\bf v}) = 0 \\
\notag\frac{\partial (\rho{\bf v})}{\partial t} &+ \nablavec \cdot (\rho{\bf vv} - {\bf BB}) + \nablavec p + \nablavec \left(\frac{B^2}{2}\right)\\ 
&= \rho{\bf g} + \nablavec \cdot \bar\bar t^{\rm v} \\
\notag\frac{\partial (\rho E)}{\partial t} &+ \nablavec \cdot \left[{\bf v}\left( \rho E + p + \frac{B^2}{2} \right) - {\bf B}({\bf v} \cdot {\bf B})\right] \\
\notag&= \rho{\bf g} \cdot {\bf v} + \nablavec \cdot ({\bf v} \cdot {\bar\bar{t}}^{\rm v} + \kappa \nablavec T)\\
&+ \nabla \cdot [{\bf B} \times (\eta\nabla \times {\bf B})]\\
\notag\frac{\partial {\bf B}}{\partial t} &+ \nabla \cdot ({\bf vB} - {\bf Bv}) \\
&= -\nabla \times (\eta\nabla \times {\bf B})
\end{align}
where
\begin{align}
E &= \frac{1}{2}v^2 + \epsilon + \frac{1}{2}\frac{B^2}{\rho}\end{align}
is the specific total energy, $\epsilon$ is the specific internal energy, $\bar\bar{t^{\rm v}}$ is the viscous stress tensor,  ${\bf g}$ is the gravitational force per unit mass, $\kappa$ is the heat conductivity, and $\eta$ is the resistivity.

FLASH implements a Direct Eulerian PPM solver \cite{woo84,col84}.  The constrained transport method \cite{eva88} is used to enforce divergence-free magnetic fields.

\subsection{Global 3D Simulations}
MHD simulations of the MRI in accretion disks are often conducted in a shearing box approximation due to the high resolution required to model MHD turbulence.  However, this approach can introduce complications and pitfalls including a limited spatial scale for the simulations, side-effects of the shearing box symmetry and artifacts from the application of periodic boundary conditions \cite{reg08} as well as an aspect ratio dependence for MRI channel solutions \cite{bod08}.  We are now at a stage where high resolution global 3D simulations are possible and this is the approach we will take.

\subsection{Non-ideal MHD}
In the disk model of Kuncic and Bicknell \cite{kun04}, jets and/or winds are primarily responsible for transporting the angular momentum necessary for accretion to proceed at a rate consistent with observations of the most powerful astrophysical sources.  We will test the hypothesis that the large-scale poloidal magnetic fields necessary for these outflows may be self-consistently generated in the accretion flow.  Recent simulations \cite{kig05} show that in the presence of an externally applied large-scale magnetic field, angular momentum transport by the vertical ($\phi z$) Maxwell stress is comparable to its radial ($r\phi$) component.  Magnetic reconnections can have a significant influence on magnetic field topologies \cite{vas75,bis93}. Notwithstanding the high degree of ionisation of plasmas in the accretion disks of AGN and X-ray binaries, we suggest that these reconnections and non-ideal MHD effects in general cannot be neglected.  Even in numerical models that do not explicitly include non-ideal effects, they can appear in the form of numerical resistivity which is difficult to control and quantify.  By explicitly modelling a finite resistivity, we will explore its effect on the evolution of the magnetic field topology, particularly the emergence of a significant $z$-component which is necessary to produce high mass accretion rates.

\subsection{Radiation}
The inclusion of radiation in our simulations is imperative for directly comparing to the observational data, which is almost exclusively in the form of photons detected in various wavebands. Most of the emission that characterises quasars and other AGN is attributed to the putative accretion disk and peaks at optical--ultraviolet spectral energies. Radiative transfer will also be required to transport the internal energy dissipated in the disk plasma at the end of a turbulent cascade, i.e. to cool the disk.  To date, no simulations have investigated the effect of optically thick radiative cooling on the MRI in full global 3D.  Including radiation will also allow us to test whether the blackbody emission from the disk is modified by outflows.  In addition, it may be that regions of the disk where radiation dominates may be thermally unstable \cite{sha76}, thus affecting the dynamics.  

Implementing radiation is computationally very demanding; numerically solving the full radiative transfer problem in 3D is currently not feasible.  Instead, a common approach is to average over frequency and solve the equations in the flux-limited diffusion (FLD) approximation \cite{alm73,lev81}, a technique which still allows one to approximate the emergent spectrum.  A ``flux limiter'' is used to interpolate between the optically thin and optically thick cases, giving a reasonable measure of the energy carried away by radiation \cite{cas04}.  FLD has previously been implemented in a shearing box in a reference frame co-moving with the fluid (e.g. \cite{tur01,hay06}).  Simulations show a stratified disk in contrast to the standard model \cite{tur04}, and that radiative diffusion dominates Poynting flux throughout the disk and the upper layers are magnetically supported and inhomogeneous, likely affecting the emergent thermal spectrum \cite{hir06,kro07,bla07}.  The implementation of MHD with FLD described in \cite{kru07} achieves energy conservation by using a mixed-frame numerical scheme, evaluating radiation quantities in the lab frame and fluid opacities in the co-moving frame, thus enabling us to address radiation in a quantitative way.  The algorithm also provides improved speed and accuracy compared to \cite{tur01,hay06}.  

Our goal is to calculate the steady-state emission spectrum of a turbulent MHD disk around a supermassive black hole in the nucleus of a galaxy and to compare the predicted spectrum with the observed optical spectra of quasars (see, e.g., \cite{kunbick07a}).

\section{Concluding Remarks}
The publication of these proceedings coincides with the 50th anniversary of the discovery of the MRI \cite{vel59}, which has had such a profound impact on our understanding not only of accretion disks -- arguably nature's most powerful energy source -- but plasmas in general, ranging from laboratory scales to galaxy scales.  Further landmarks in accretion disk theory came with the laying down of the standard theory \cite{sha73} in what remains the most cited paper in all of astrophysics; the discovery that large scale magnetic fields can vertically transport matter, energy and momentum \cite{bla82}; the first numerical MHD simulations \cite{uch85,shi86}; as well as the rediscovery of the MRI in accretion disks and the demonstration by numerical simulations that this is indeed the MHD process anticipated by the analytic standard theory necessary to produce the turbulent radial transport of energy and momentum  \cite{bal91,bal98}.  Despite these major steps forward in assembling the components of a complete accretion disk theory, we have not to date seen numerical simulations which can \emph{self-consistently} produce all the salient features of quasars and other high energy astrophysical sources: high mass accretion rates, outflows, winds and jets, the formation of a magnetised corona, and the observed thermal spectrum.

We expect it will again be magnetic fields which will hold the key to resolving these outstanding issues.  To this end, we are developing new 3D global MHD simulations to test the hypothesis that small-scale, stochastic fields can self-consistently generate the large-scale poloidal magnetic fields necessary for the transport of energy and momentum from accreting matter necessary to produce each of the above observed features.  The analytical basis for this generalized model was laid down by the companion work \cite{kun04} to our current numerical modelling, which the rapid advances in computing power accompanied by new codes and more efficient algorithms has now made possible.

It is an exciting time for accretion disk theory and for our understanding of the microphysics that drives these and other plasma systems in nature.

\end{document}